\documentclass[aps,prb,showpacs,twocolumn]{revtex4}  

\usepackage{graphicx}
\usepackage{amssymb}                

\def\newblock{\hskip .11em plus .33em minus .07em}
\def\gapx{\lower 2pt \hbox{$\buildrel>\over{\scriptstyle{\sim}}$\ }}
\def\lapx{\lower 2pt \hbox{$\buildrel<\over{\scriptstyle{\sim}}$\ }}
\def\he4{$^4$He}
\def\paraH2{{\it p}-H$_2$}
\def\orthoD2{{\it o}-D$_2$}
\def\Am2{\AA$^{-2}$}

\begin{document}

\title[Disorder and the elusive superfluid phase of \textit{para}-hydrogen]{Disorder and the elusive superfluid phase of \textit{para}-hydrogen}
\author{Joseph Turnbull and Massimo Boninsegni} 
\affiliation{Department of Physics, University of Alberta, Edmonton, 
    Alberta, Canada T6G 2J1}
\date{\today}

\begin{abstract} 

The possibility that disorder may stabilize a superfluid phase of {\it para}-hydrogen  in two dimensions is investigated theoretically by means of Quantum Monte Carlo simulations. We model disorder using a random distribution of scatterers, and study the thermodynamic behavior of the system as a function of the scatterer density. Disorder gives rise to equilibrium glassy phases with no superfluid properties. Indeed, the propensity for quantum exchanges of hydrogen molecules is reduced even with respect to what is observed if the scatterers are arranged as a regular crystal, a physical setting that also yields no superfluidity.

\end{abstract}

\maketitle

\section{INTRODUCTION}

The inability of observing experimentally  a superfluid phase of bulk {\it para}-hydrogen (\paraH2),\cite{ginzburg72} is due to the crystallization of the system at a temperature $T_{\rm c} \approx$ 14 K, significantly higher than that ($T$ $\approx$ 6 K) at which phenomena such as Bose-Einstein condensation (BEC) and presumably superfluidity (SF), may be expected to occur in the liquid phase.  Crystallization takes place because of the depth of the attractive well of the potential between two hydrogen molecules, approximately three times that between two helium atoms. 

The macroscopic property of SF is theoretically associated with the onset of long  exchange cycles,  involving macroscopic numbers of identical particles.  As quantum exchanges are greatly suppressed in the crystal phase, due to particle localization, it is widely accepted that the observation of SF of \paraH2 hinges on the stabilization of a low-temperature liquid phase (although solid order may not be entirely incompatible with superflow.\cite{Penrose56, Anderson84, Lifshitz69, Chester70, Leggett70, Chan04})
Several attempts have been made\cite{bretz81,maris86,maris87,schindler96} to supercool bulk liquid \paraH2, but the experimental hurdles have thus far made it unfeasible to reach a temperature in the range where SF might be observed. 

Other experimental avenues have been explored, aimed at suppressing crystallization. They include confinement and reduction of dimensionality (which reduces the coordination number).  For example, the phase diagram and structure of monolayer \paraH2 films adsorbed on graphite have been studied by various techniques.\cite {nielsen80,lauter90,wiechert91,vilches92}

One of the most remarkable aspects\cite{vilches92} is that the melting temperature $T_m$ of a solid \paraH2 monolayer can be significantly less than that of bulk \paraH2.  Yet, there appears to be a limit to what can be achieved by reduction of dimensionality alone. A theoretical study\cite{boninsegni04b} of the phase diagram of \paraH2 in purely two dimensions (2D) has shown that the equilibrium phase of the system at low $T$ is a triangular crystal, with no evidence of even a metastable liquid phase at lower density. Indeed, the system remains  solid  all the way to the spinodal density.  The melting temperature $T_{\rm m}\sim$ of the equilibrium crystal is 6.8 K, i.e., approximately half that of bulk \paraH2 but still significantly higher than the temperature at which the system, if it remained a liquid, would turn superfluid, estimated at $\sim$ 2 K in 2D. Later studies\cite{boninsegni04_li, turnbull07_d2} of \paraH2 adsorbed onto weakly attractive substrates, predicted only a very slight reduction of $T_{\rm m}$ with respect to the purely 2D case, due to an enhancement of the kinetic energy arising from zero-point motion in the direction perpendicular to the substrate (for example, in the case of lithium, arguably one of the weakest substrates existing, a reduction of merely 0.3 K was observed\cite{boninsegni04_li}).  This suggests that mere reduction of dimensionality is not sufficient to stabilize a fluid phase of \paraH2 down to low enough a temperature for superfluidity to be observed.

In 1997, Gordillo and Ceperley\cite{Ceperley97} (GC) proposed that a significant reduction of the equilibrium density of  2D  \paraH2, and the ensuing stabilization of a low temperature liquid phase,  could be achieved by intercalating a \paraH2 fluid within a crystal of impurities,  incommensurate with the equilibrium crystal structure of pure 2D \paraH2. Path Integral Monte Carlo (PIMC) simulations of such a model system, yielded some evidence of a possible superfluid transition at $T \sim$ 1 K.  However, the size of the simulated system was very small  (4 impurities and of order 10 \paraH2 molecules).  A subsequent PIMC study,\cite{boninsegni_alk} comprising up to ten times as many particles,  strongly suggests that {\it a}) the equilibrium phase of \paraH2 in such a setting is not liquid, but instead exhibits long range crystalline order, and {\it b}) the observation of superfluidity in Ref.   \onlinecite{Ceperley97} is merely a finite-size effect (this latter finding is confirmed here).

In this work, we explore theoretically the scenario of a 2D \paraH2 fluid in a disordered environment. Our aim is again to determine whether a non-crystalline equilibrium phase, displaying long-range superfluid coherence, may arise as a result of the frustration of  solid long-range order induced by the random potential.  That disorder should promote SF is certainly not obvious, and may even be counter-intuitive. In fact, disorder is known to promote {\it  localization}, with the ensuing {\it disappearance} of SF in a system of hard core bosons.\cite{fisher89} However, recent numerical work has yielded evidence of a possible ``superfluid glassy" phase of condensed helium, characterized by simultaneous broken translational invariance (and therefore nonzero shear modulus) {\it and} SF, with no diagonal long-range order.\cite{MBsuperglass} It is conceivable that a similar phase of \paraH2 could arise in disorder.

We investigate the above scenario by means of  Quantum Monte Carlo simulations based on the continuous-space Worm Algorithm (WA). Specifically,  we simulate a fluid of \paraH2 molecules moving in 2D in the presence of a static potential generated by a random distribution of identical impurities, averaging the results over several independent realizations of the external potential. We have assumed a specific simple model potential,  so as to describe the interaction between each scatterer and the \paraH2 molecules, and varied the overall strength of the disordering potential by changing the density of impurities.  The main result of this study is a {\it null} one, i.e., even though we observe  exchanges of small groups of molecules, they have a local character, i.e., long permutation cycles spanning the entire system do not occur; as a result, a finite superfluid signal never materializes in the low temperature and thermodynamic limits.  We observe a glassy phase, with broken translational invariance, but we also clearly observe that \paraH2 molecules {\it locally} recreate the triangular lattice structure associated with bulk 2D \paraH2,\cite{boninsegni04b} interrupted by the underlying impurity matrix.  

Based on this finding, as well as on the null results yielded by various simulation works carried out over the past few years, we propose that a superfluid phase of 2D \paraH2 will {\it not} be observed, in the absence of some indirect physical mechanism capable of  renormalizing (weakening) the effective interaction among molecules.

The remainder of this manuscript is organized as follows:  Section \ref{model} offers a description of the model used for our systems of interest, including a discussion of the potentials and the justifications for the main underlying  assumptions.  Section \ref{method} involves a brief discussion of the computational techniques employed, in addition to details of calibration and optimization.  The results are presented in Sections \ref{results1}, \ref{results2}, and \ref{results3}; finally, Section \ref{conclusions} is a summary of the findings and our concluding remarks.

\section{MODEL}
\label{model}

We consider a system of $N$ identical hydrogen molecules (composite bosons) in the \textit{para} nuclear spin state, intercalated within a strictly 2D array of identical impurities, the entire system enclosed within a simulation cell of sides $L_{x}$ and $L_{y}$ ($\mathcal A$=$L_{x} \times L_{y}$), with periodic boundary conditions in both directions.  All of the ($M$) impurities and hydrogen molecules are regarded as point particles,  with impurities fixed in space at positions ${\bf R}_k,\ k=1,...,M$.   The model quantum many-body Hamiltonian is therefore as follows: 

\begin{eqnarray}\label{hm}
\hat{H}&=&-\frac{\hbar^{2}}{2m}\sum_{i=1}^{N}\nabla_{i}^{2}  +
\sum_{i<j}V(r_{ij})  + \sum_{i=1}^N\sum_{k=1}^{M} U (|{\bf r}_i-{\bf R}_k|)
\end{eqnarray}

Here, $m$ is the mass of a hydrogen molecule, $\{{\bf r}_j\}$ (with $j$=1,2,...,$N$) are the positions of the hydrogen molecules, and $r_{ij}\equiv |{\bf r}_i-{\bf r}_j|$. $V$ is the potential describing the interaction between any two hydrogen molecules, and $U$ represents the interaction of a hydrogen molecule with an impurity.  All pair potentials are assumed to depend only on relative distances.  Most of the results presented here apply to the case where the positions of the impurities are random, but obviously the above microscopic model can describe regular, crystalline impurity arrays (indeed, the very same model was adopted for that purpose in Refs.   \onlinecite{Ceperley97} and \onlinecite{boninsegni_alk}).

The interaction  $V$ is described by the Silvera-Goldman potential,\cite{Silvera1} which provides an accurate description of energetic and structural properties of condensed \paraH2.\cite{johnson96,operetto}   The interaction of a hydrogen molecule with each  impurity scatterer is modeled using a standard 6-12 Lennard-Jones (LJ) potential, with parameters $\epsilon=9.54$ K and $\sigma=3.75$ \AA\, as utilized in Ref. \onlinecite{Ceperley97}. We wish to stress that, whether or not such a model potential offers an accurate description of any realistic interaction is largely immaterial for our study, as our aim is to study the general effect of disorder; in fact, in principle just about {\it any} interaction could be utilized, if disorder is produced by randomly placed scatterers of variable density.

In addition to restricting the dimensionality of the system to $d$=2, the model (\ref{hm})  clearly contains important physical simplifications, such as the restriction to additive pairwise interactions (to the exclusion of, for example, three-body terms), all taken to be central.  However, here we are primarily interested in examining the fundamental physics contained in a general many-body problem described by a Hamiltonian such as (\ref{hm}), rather than making quantitative predictions about an actual experimental  system - though it should be noted that an actual experimental realization may be feasible.\cite{Cole92,Leatherman96}

\section{Computational Method}  
\label{method}

Thermal expectation values for quantum many-body systems described by a Hamiltonian such as (\ref{hm}) can be computed by means of Quantum Monte Carlo (QMC) simulations. This technique provides numerical estimates that one may regard for all practical purposes as as {\it exact} (at least or Bose systems, of interest here), as results are only affected by a {\it statistical} uncertainty that can be rendered negligible in most cases. In this study, finite-temperature estimates were obtained making use of the continuous-space \textit{worm algorithm}, described in Refs. \onlinecite{boninsegni06worm1} and  \onlinecite{boninsegni06worm2}.

The only inputs in the simulations are the Hamiltonian and the temperature.  As with standard PIMC, the calculation has a finite time step error, which can be made sufficiently small by taking the time step short enough; for all of the system sizes explored, it was found that convergence of the energy estimate is achieved (within statistical uncertainty) using a value of the time step $\tau_{w}= 1/640$ K$^{-1}$.   Convergence for properties such as the superfluid density is reached using a much longer time-step, and this property has been taken advantage of, where indicated.  Based on comparisons of results obtained from simulations with different values of the time step, we estimate our systematic error on the total energy per \paraH2 molecule (the observable that is most sensitive to time-step error) to be of the order of 0.1 K or less (of order 0.6\%).  

\subsection{Simulations} \label{sims}

All of the simulation results presented below are obtained starting from an initial configuration of hydrogen molecules randomly dispersed within the impurity array.  We have studied the physical behavior of the system at sufficiently low temperature, with the aim of extrapolating the results to the $T$=0 limit. Specifically, the lowest temperature for which estimates were obtained is 3.125$\times$10$^{-2}$ K. 

Our simulations for \paraH2 in a disordered environment are carried out using two different simulation cells. The first is a 30$\times$30 \AA$^{2}$ square cell, wherein we arrange $M$=12, 16, and 20 impurities (corresponding to impurity densities $\Omega$=0.0133 \Am2, 0.01778 \Am2, and 0.0222 \Am2 respectively) in a completely random fashion.  The second cell, with dimensions 40$\times$34.64 \AA$^{2}$, contains $M$=16 randomly placed impurities (corresponding to $\Omega$=0.01155 \Am2). For each impurity density, eight different realizations are considered (i.e., different random positions of the impurities).  

In a slightly different arrangement, we have considered a 40$\times$34.64 \AA$^{2}$ cell, in which we first arranged $M$=16 impurities on a regular triangular lattice (details are given below), and then applied to each impurity are  random displacements $D_{0,1}\in$(0.0,1.0) \AA, $D_{0.5,2}\in$(0.5,2.0) \AA, $D_{0,2}\in$(0.0,2.0) \AA, and $D_{1,2}\in$(1.0,2.0) \AA. Eight such ``nearly-periodic" impurity matrices were realized in each case (32 total).  For each realization, calculations for a range of \paraH2 coverages were carried out for selected coverages.  

For each realization and each set of system parameters, energetic and structural properties were studied; in general, these remain unchanged below $T$=2 K. Superfluid properties of the system are investigated through the direct computation of the superfluid density using the traditional winding number estimator;\cite{Ceperley1} we as well obtain the one-body density matrix, which is readily accessible using the worm algorithm.\cite{boninsegni06worm2}

A detailed discussion of the results of these simulations is provided in Sections \ref{results2} and \ref{results3}. However, in order to provide a useful baseline for comparison, we shall first discuss our results  for a {\it regular}, crystalline matrix of impurities arranged on a triangular lattice, i.e.,  the same system already investigated  in Refs. \onlinecite{Ceperley97} and \onlinecite{boninsegni_alk}. 

\section{Results: \paraH2 intercalated within a periodic 2D impurity matrix}  \label{results1}

We have obtained results for the same three system sizes of Ref. \onlinecite{boninsegni_alk}, i.e., the simulation cell contains triangular lattices consisting of 2$\times$2 (cell area is ${\cal A}$ = 20$\times$17.32 \AA$^{2}$), 4$\times$4 (${\cal A}$ = 40$\times$34.64 \AA$^{2}$), and 5$\times$6 (${\cal A}$ = 50$\times$51.96 \AA$^{2}$) impurity atoms with 10.0 \AA\ nearest neighbor spacing. 
In order to assess the robustness of our predictions in the thermodynamic limit, we have also performed  simulations for  a 10$\times$12 (100$\times$103.92 \AA$^{2}$) lattice of impurities, at $T$=4 K and $T$=0.25 K (the latter using a 4$\times\;$ larger time step).  We use these results largely to infer/confirm structural properties.

\begin{figure}             
\centerline{\includegraphics[width=3.00in]{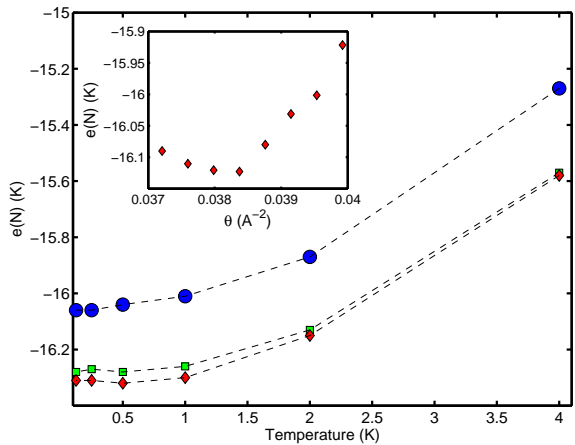}} 
\caption{(Color online) Energy per \paraH2 molecule for a regular crystalline arrangement of impurities, at the equilibrium coverage $\theta_e$=0.0381 \Am2 as a function of the temperature, for different system sizes. Different symbols show estimates for lattices of 4 (circles), 16 (squares) and 30 (diamonds) impurities. Inset shows the energy per molecule computed at $T$=1 K as a function of the coverage $\theta$, obtained for a system of 30 impurities.}
\label{ene}
\end{figure}

In order to obtain the ground state equation of state of the system, we have computed the energy per particle $e$ as a function of coverage $\theta$, extrapolating the results to the $T\to 0$ limit. In all energy calculations, corrections from the contribution to the potential energy arising from particles (and their periodic images) beyond the interaction cutoff radii were incorporated. Our results (Fig. \ref{ene}) are identical, within statistical uncertainties, with those of Ref. \onlinecite{boninsegni_alk}. In particular, no significant dependence of the energy on temperature is observed for $T$ below 2 K, and even the dependence on the size of the system  (for all but the smallest one considered) is within, or very close to, statistical error.  Our results are also in {\it qualitative} agreement with those of Ref. \onlinecite{Ceperley97}, although estimates offered therein are shifted upwardly by as much as a few K (the reasons for this discrepancy remain unclear, as merely finite size corrections can not account for the difference).

The equilibrium 2D density $\theta_{e}$ is 0.0381 \Am2, which, as we shall show below, corresponds to a {\it commensurate} superstructure formed by the \paraH2 molecules. Furthermore, a stability analysis of the data for $e(\theta)$ (shown in the inset of Fig. \ref{ene}) indicates that doping above $\theta_e$ will {\it not} result in a homogeneous phase, but rather in the {\it coexistence} of  two (commensurate) phases.

\begin{figure*}             
\centerline{\includegraphics[width=6.00in]{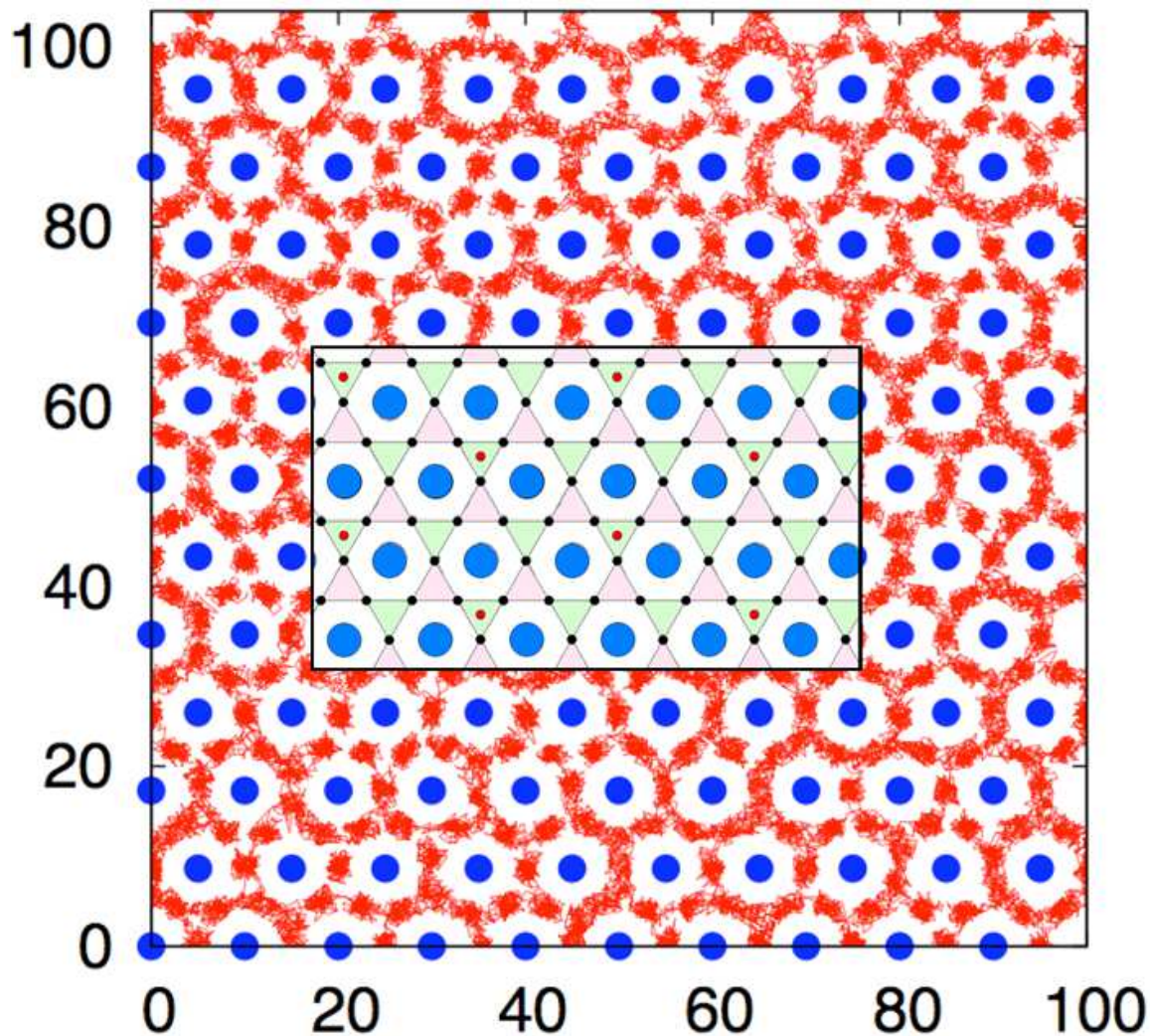}} 
\caption{(Color online) Snapshot of a typical configuration of \paraH2 molecules at low $T$ at $\theta=\theta_e$, intercalated within the periodic 2D fixed impurity matrix (described in the text).  The positions of all \paraH2 molecules at each one of the imaginary time slices are shown as discrete paths, while impurities are shown as solid circles.  All distances are in \AA.  Inset shows a schematic representation of the ensuing phase, inferred from a visual inspection of configurations. Large  circles represent impurities, whereas smaller circles are \paraH2 molecules. Red circles are \paraH2 molecules forming the 1/3 commensurate triangular phase, which can occupy either one of two equivalent lattices (different color shades).}
\label{huge snapshot}
\end{figure*} 

Figure \ref{huge snapshot} shows a snapshot of the \paraH2 world lines for a system with equilibrium \paraH2 coverage $\theta$=0.0381 \Am2 at $T=$ 0.25 K on the 10$\times$12 cell (corresponding to 400 \paraH2 molecules, precisely 3$+\frac{1}{3}$ times the impurity density). The positions of all \paraH2 molecules at each one of the imaginary-time slices are shown as discrete paths, while the fixed impurities are shown as solid circles.  The spread of the imaginary-time path for each \paraH2 molecule gives a measure of their quantum delocalization (zero-point motion).  

One can immediately  identify a trihexagonal tiling of \paraH2 corresponding to a kagom\'{e} lattice commensurate with, and having three times the density of the impurity background (i.e. three \paraH2 per centroid of every nearest-neighbor impurity triplet).  Superimposed over this structure is an additional 1/3 coverage of \paraH2 molecules arranged on a triangular lattice; sites of this triangular lattice correspond to regions of the total structure having four \paraH2 molecules per centroid of nearest-neighbor impurity triplets (schematically shown in inset of Fig. \ref{huge snapshot}). As mentioned above it is found, using the procedure described in Ref. \onlinecite{turnbull07_d2}, that this phase is incompressible.  

Fundamental theoretical arguments,\cite{proko2005} as well as simulation of hard core bosons on the triangular lattice\cite{bontr} (with which the commensurate phase of \paraH2 described above bears significant similarities and seems a close physical realization) strongly suggest that the commensurate phase of \paraH2 found here is not {\it supersolid}. 
In any case, in order to assess the superfluid response of the system, one can compute the superfluid density, $\rho_s$,  as a function of temperature. Accurate scaling of numerical results obtained for systems of sufficiently different and large sizes is {\it of paramount importance} in the presence of an underlying crystal, with the ensuing loss of translational invariance,  in order to afford a reliable extrapolation  of the physical behavior characterizing the thermodynamic limit. The system of interest here offers a chief example of  the importance of this aspect.

For example, the estimates of $\rho_s$ for the 4$\times$4 impurity lattice, which are in agreement with the results obtained for the 2$\times$2 system, yield a finite value of $\rho_s$, approximately equal to 25\%.  However, upon increasing the system size to the 5$\times$6 cell, this finite superfluid signal goes down to a mere 5\%, vanishing altogether  for the 10$\times$12 cell; concurrently,  the one-body density matrix  displays a clear exponential decay with distance, when evaluated on sufficiently large size systems. It should be noted that the change of behavior observed on extending the system size is {\it not} attributable to some loss of efficiency of the simulation method in sampling long permutations of \paraH2 molecules. While this would be certainly an issue with conventional PIMC, the continuous-space Worm Algorithm does not suffer from any such system size limitation, as shown by a number of calculations carried out over the past few years, e.g., for extended defects in solid helium, including several thousand particles.\cite{pollet07,boninsegni07}

Rather, a large size is simply required in order for the actual, non-trivial equilibrium structure of the system to be possible, as shown in Fig. \ref{huge snapshot}. Our analysis reveals that \paraH2 forms a commensurate crystal phase with no interstitials or defects.  Such a phase, masked for smaller system sizes (for which, while the obtained structure is artificially frustrated, remnants of the crystalline phase are noticeable only {\it a posteriori}), is insulating,\cite{proko2005} with explicit calculations of the superfluid density lending confirmation.  Thus, in agreement with Ref. \onlinecite{boninsegni_alk}, we conclude that all previously reported superfluid properties are finite-size artifacts, and that \paraH2 in this geometry is necessarily a regular quantum solid in the thermodynamic limit.

This does not mean that exchanges of \paraH2 molecules do {\it not} take place; however, they are largely local in character; long cycles spanning the whole system and resulting in a finite superfluid response, are exponentially suppressed in the thermodynamic limit.

\section{Results: \paraH2 intercalated within a random 2D impurity matrix}  \label{results2}

We now turn to describing the results obtained for {\it completely random} placement of scatterers.
For each of the eight independent realizations of a random distribution of scatterers at a given density, we compute the ground state equilibrium \paraH2 coverage $\theta_c$; in general, the value of $\theta_c$ depends on the specific impurity realization, particularly for smaller system sizes. It is found to fluctuate typically within $\sim$10\% from one realization to another. This gives an idea of the local density fluctuations that occur in the thermodynamic limit, as a result of the inhomogeneity of the system.

\begin{figure}             
\centerline {\includegraphics[width=3.00in]{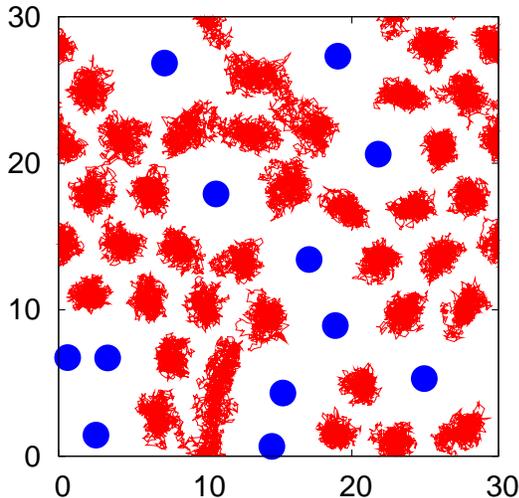}} 
\caption{(Color online) Snapshot of a typical configuration of \paraH2 molecules at $T$= 0.5 K for $\theta$=$\theta_{e}$, intercalated within the 2D matrix of randomly placed impurities with $\Omega$=0.0133 \Am2.  The positions of all \paraH2 molecules at each one of the imaginary time slices are shown as discrete paths, while impurities are shown as solid circles.  All distances are expressed in \AA.}
\label{snapshots0}
\end{figure} 

The extrapolated ground state energy per particle, $e(N)$, for a disordered arrangement of impurities, roughly  interpolates between that for bulk 2D \paraH2  (see, for instance, Ref. \onlinecite{boninsegni04b}) and the energy for \paraH2 in the presence of periodic impurity arrays of the same density,\cite{boninsegni_alk} as should be expected.

Fig. \ref{snapshots0} shows a snapshot of the \paraH2 world lines for a system at the equilibrium coverage $\theta_{e}\approx$ 0.052 \Am2, for a specific realization of the disorder with an impurity density $\Omega=0.0133$ \Am2 (the temperature is $T=$ 0.5 K). By visual inspection, we can clearly establish  that the \paraH2 molecules attempt to recreate the triangular lattice associated with bulk 2D \paraH2,\cite{boninsegni04b} interrupted by the underlying impurity matrix. This is consistently observed for  all realizations of the disorder, at all impurity densities. Moreover, the observed little or no overlap among quantum-mechanical delocalization ``clouds" of the different molecules are consistent with quantum exchanges being suppressed in this system. 

\begin{figure}            
\centerline{\includegraphics[width=3.00in]{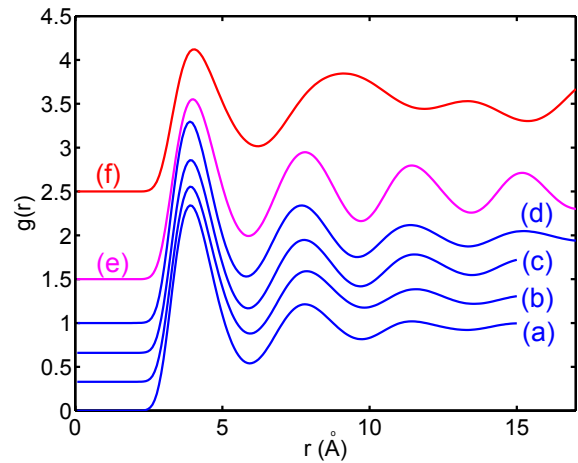}}
\caption{(Color online) Ground state \paraH2 pair correlation functions ($g(r)$) for equilibrium \paraH2 coverage.   All functions have been uniformly displaced upwards by their y-intercept to ease comparison. Results {\it a} through {\it d} refer to randomly placed impurities with decreasing density (from $\Omega$=0.022 ({\it a}) to 0.01155 ({\it d}) \Am2). Also shown for comparison are $g(r)$ for bulk 2D \paraH2 ({\it e}), as well as for the case of a triangular impurity lattice ({\it f}), with impurity density $\Omega$=0.01155 \Am2. }   
\label{profile2}
\end{figure} 

The above findings are confirmed by an examination of the \paraH2 pair correlation function, shown in Figure \ref{profile2} for disordered matrices of different impurity densities. Also shown for comparison is the pair correlation function for bulk 2D \paraH2 at the equilibrium coverage ($\theta$=0.056 \Am2, from Ref. \onlinecite{boninsegni04b}), as well as for the case of a regular impurity matrix of density $\Omega$=0.01155 \Am2. The progressive loss of structure at long distance, as the concentration of impurities is increased, is obvious; in contrast, the short-range features of the pair correlation function are essentially unaffected, largely reproducing those of the pair correlation function for bulk 2D \paraH2. 

Analysis of the one-body density matrix reveals a rapid exponential decay, with no signature of a Kosterlitz-Thouless power-law decay.  For each realization, we calculate a superfluid density of precisely zero, with quantum exchanges between \paraH2 molecules being strongly suppressed with respect to the systems with periodic impurity matrices; system-spanning permutations, necessary for the system  to feature a finite superfluid response, are never generated.  All evidence points to this system being a ``Bose glass", devoid of both solid order and off-diagonal quasi-long-range order. Clearly, no enhancement of Bose statistics is brought about by disorder. In order to gain additional insight into the effect of disorder, we next study a physical arrangement of impurities that interpolates between the regular periodic array of  Section \ref{results1} and the disordered configuration of Section \ref{results2}.  

\section{Results: \paraH2 intercalated within a nearly-periodic 2D impurity matrix}  \label{results3}

\begin{figure}             
\centerline{\includegraphics[width=3.00in]{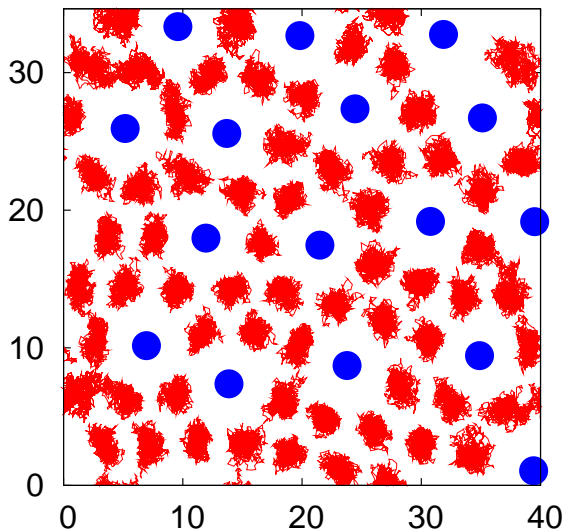}} 
\caption{(Color online) Snapshot of a typical configuration of \paraH2 molecules at $T=$ 0.5 K for $\theta$=$\theta_{e}$, intercalated within the aperiodic 2D fixed impurity matrix with density $\Omega$=0.01155 \Am2.  The positions of all \paraH2 molecules at each one of the imaginary time slices are shown as discrete paths, while impurities are shown as solid circles.  All distances are expressed in \AA.  }
\label{snapshots2}
\end{figure} 

For the case where impurities are located at positions obtained applying random displacements $D_{1,2}$  to  regular triangular lattice sites (see Sec. \ref{sims}),  we find on average a 13\% enhancement of the equilibrium \paraH2 coverage obtained in Section \ref{results1}.

Structurally, one may infer from the pair correlation function (f) in Figure \ref{profile2} that the \paraH2 is in a liquid-like or glassy state. Shown in Figure \ref{snapshots2} is a typical world-line snapshot at $T$=0.5 K.  Upon examination of many such ``snapshots", we confirm that the (very limited) interparticle overlap observed in Figure \ref{snapshots2} is the norm.  

An accumulation of statistics indicate an enormous suppression of quantum exchanges, almost entirely absent.  In every disordered impurity realization examined, the superfluid fraction $\rho_s$ is found to be exactly zero.  The structure and lack of superflow exhibited when $D_{1,2}$ 
random displacements are applied, are also observed  for $D_{0.5,2}$ and $D_{0,2}$ 
random displacements, with no significant physical changes observed in these three cases across all realizations.  Altogether, these degrees of disorder imparted to the system are sufficient to induce strong particle localization, and eliminate even the local exchanges found for periodic arrays of impurities in smaller systems, as is the case for the systems studied in Section \ref{results2}.

Upon reducing the magnitude of the impurity displacements to $D_{0,1}$ \AA, we find significant variation from one realization to the next, with between a 0\% and 4\% enhancement of the equilibrium \paraH2 coverage with respect to the system having a periodic array of impurities (corresponding to at most 2 additional interstitial particles).

In realizations for which the particle density is unchanged, the same insulating phase found in Section \ref{results1} is recovered, albeit with a very slight broadening of $g(r)$.  

In the realizations where additional molecules happen to be introduced (for example, in realizations where three impurities that are part of one nearest-neighbor triplet are displaced radially outward from their centroid by order 1.0 \AA), $\rho_s$ is zero for all $T$, and the broadening of $g(r)$ with respect to that of Section \ref{results1} is significant.  

These results are adequate to conclude that the introduction of disorder which is strong enough to introduce additional particles into the \paraH2 crystal leads to acute particle localization and a suppression of exchanges; disorder weak enough to not change the equilibrium coverage does not cause a perturbation in the necessarily insulating \paraH2 crystal structure.

\section{CONCLUSIONS}
\label{conclusions}

Using numerically exact finite-temperature path integral Monte Carlo (worm algorithm) methods, we studied purely 2D \paraH2 intercalated within an array of impurities.  We performed calculations based a simple model, in which the impurities are assumed static and point-like, and \paraH2-impurity interactions are given by a Lennard-Jones potential.  

For \paraH2 intercalated within a periodic array of scatterers, the explicitly observed crystalline structure of hydrogen allows us to conclude with confidence that the results seen for the largest system sizes are indicative of the physics in the thermodynamic limit, and not a manifestation of algorithmic deficiency.  The finite superfluid response observed in earlier simulations\cite{Ceperley97} of small systems (where the crystal structure could not be exactly realized) are thus attributable to finite-size effects.

From our simulations of \paraH2 in the presence of disordered arrays of impurities, we conclude that the introduction of such disorder enhances localization of \paraH2 molecules, suppressing quantum exchanges.  Though disorder can give rise to a glassy phase, exchange frequencies are \textit{suppressed} with respect to the exchange statistics observed in the systems with periodic arrays of impurities, in agreement with the conventional view of disorder-induced (Anderson) localization in strongly-interacting Bose systems.  

Our primary conclusion is that \paraH2 intercalated within a 2D periodic or aperiodic array of impurities is not a candidate for the observation of superfluidity.  

The prognosis, it seems, for observing superfluidity of \paraH2 in 2D films, is not an optimistic one.  Though it is possible that incommensuration may permit an enhancement of disorder and even the formation of a fluid-like state, as shown in recent studies of \paraH2 and \orthoD2 adsorbed on a substrate composed of graphite pre-plated by a single atomic layer of krypton,\cite{wiechert04,turnbull07_d2} it is doubtful that this particular flavor of disorder can enhance particle exchange, and indeed here appears to instead suppress it as well.  Based on this collection of null results, we conclude that a superfluid phase of 2D \paraH2 is {\it not} likely to be observed, in the absence of some indirect physical mechanism capable of  renormalizing (weakening) the effective interaction among molecules.

Barring such a mechanism, superfluid behavior in  \paraH2 may be observed experimentally in finite size systems only, e.g., clusters.\cite{mez1,Fabio08a}

\section*{Acknowledgments}
This work was supported by the Natural Sciences and Engineering Research Council of Canada (NSERC) under research Grant No. G121210893 and by the Alberta Informatics Circle of Research Excellence (ICORE). Computing support from Westgrid is gratefully acknowledged.

\end{document}